\documentclass[12pt,preprint]{aastex}

\def\hho  {H$_2$O}
\def\kms  {\ifmmode {{\rm km~s}^{-1}} \else {km~s$^{-1}$} \fi}
\def\kmsperyr  {\ifmmode {{\rm km~s}^{-1} {\rm yr}^{-1}} 
                \else {km~s$^{-1}$ yr$^{-1}$} \fi}

\def\etal {et al.~}
\def\eg   {e.g.~}
\def\ie   {i.e.~}
\def\UGC  {UGC~3789}
\def\NGC  {NGC~4258}
\def\Vlsr {\ifmmode {V_{\rm LSR}} \else {$V_{\rm LSR}$} \fi}
\def\Vhelio {\ifmmode {V_{\rm Helio}} \else {$V_{\rm Helio}$} \fi}
\def\Ho   {\ifmmode {H_0} \else {$H_0$} \fi}
\def\Msun {\ifmmode {M_\odot} \else {$M_\odot$} \fi}
\def\uvdata{({\it u,v})-data}

\slugcomment{Version: 2008 November 21}
\shorttitle{VLBI observations of \UGC} 
\shortauthors{Reid \etal}

\begin{document}

\title{The Megamaser Cosmology Project: I. \\
       VLBI observations of \UGC}

\author{M. J. Reid\altaffilmark{1}, J. A. Braatz\altaffilmark{2},
        J. J. Condon\altaffilmark{2}, L. J. Greenhill\altaffilmark{1}, 
        C. Henkel\altaffilmark{3}, K. Y. Lo\altaffilmark{2} }

\altaffiltext{1}{Harvard-Smithsonian Center for
   Astrophysics, 60 Garden Street, Cambridge, MA 02138, USA}
\altaffiltext{2}{National Radio Astronomy Observatory,
   520 Edgemont Road, Charlottesville, VA 22903}
\altaffiltext{3}{Max-Planck-Institut f\"ur Radioastronomie, 
   Auf dem H\"ugel 69, 53121 Bonn, Germany}

\begin{abstract}
The Megamaser Cosmology Project (MCP) seeks to measure the Hubble Constant
(\Ho) in order to improve the extragalactic distance scale and constrain 
the nature of dark energy.  We are searching
for sources of \hho\ maser emission from AGN with sub-pc accretion 
disks, as in \NGC, and following up these discoveries with Very Long 
Baseline Interferometric (VLBI) imaging and spectral monitoring.
Here we present a VLBI map of the \hho\ masers toward \UGC, 
a galaxy well into the Hubble Flow.  We have observed masers moving at 
rotational speeds up to $800$~\kms\ at radii as small as $0.08$~pc.
Our map reveals masers in a nearly edge-on disk in Keplerian rotation 
about a $10^7$~\Msun\ supermassive black hole. 
When combined with centripetal accelerations, obtained by observing spectral
drifts of maser features (to be presented in Paper II), the \UGC\ masers 
may provide an accurate determination of \Ho, independent of luminosities 
and metallicity and extinction corrections.
\end{abstract}


\section{Introduction}

The current ``concordance'' cosmological model assumes a flat $\Lambda$CDM
universe composed of baryons, cold dark matter, and ``dark
energy'' that accelerates the expansion of the universe \citep{Spergel:03}.
The location of the first peak in the angular power spectrum of the 
cosmic microwave background (CMB) radiation determines the 
angular-size distance to the surface of last scattering.  This
distance depends on the amount of dark energy and the geometry 
and current expansion rate of the universe, \Ho.  If one does
not assume that the universe is flat, the CMB data alone are consistent 
with a wide range of values of \Ho.  Thus, independent measurements 
of \Ho\ are needed to justify the flatness assumption and to determine 
whether dark energy is the cosmological constant, $\Lambda$, of General 
Relativity, a variable ``quintessence" \citep{Wetterich:88,Ratra:88}, 
or possibly something else. \citet{Hu:05} concludes that the most 
important single complement to CMB data would be a precise 
(\eg\ $\sim1$\% uncertainty) measurement of \Ho.

The current ``best value'' for the Hubble Constant,
$\Ho = 72 \pm 7~\kms~{\rm Mpc}^{-1}$ from the HST
Key Project \citep{Freedman:01}, is based on luminosity distance
measurements to extragalactic Cepheid variables treated as 
``standard candles.''  The 10\% uncertainty in \Ho\ is dominated by 
systematic errors that cannot easily be reduced by observations of 
more galaxies.

Very Long Baseline Array (VLBA) observations of the \hho\ megamaser 
in the nearby Seyfert 2 galaxy \NGC\  have provided an accurate, 
angular-diameter  distance to the galaxy \citep{Herrnstein:99}, 
bypassing the problems of standard candles.  The \hho\ masers in \NGC\  
arise in a thin (annular) disk viewed nearly edge-on \citep{Greenhill:95a}
and appear at galactocentric radii $R \approx 0.14~{\rm to}~0.28$~pc.
Maser lines near the systemic velocity of the galaxy come from gas moving
across the sky on the near side of the disk, and ``high-velocity lines,'' 
with relative velocities of up to $V \approx \pm1100$~\kms, come from gas 
moving along the line of sight at the disk tangent points.  
The high-velocity lines display a Keplerian rotation curve, implying a 
central mass of $\approx4 \times 10^7$~\Msun, presumably in the
form of a supermassive black hole (SMBH) \citep{Miyoshi:95}.
  
For \NGC, the velocities of individual systemic features are 
observed to increase by $\approx 9~\kms~{\rm yr}^{-1}$ 
\citep{Haschick:94,Greenhill:95b},
allowing a direct measurement of the centripetal acceleration 
($a = V^2/R$) of clouds moving across our line of sight near the 
nucleus \citep{Watson:94}.  
Conceptually, the angular-diameter distance, $D_\theta$, to \NGC\  
can be determined geometrically by dividing the {\it linear} radius of masers, 
measured from Doppler shifts and accelerations ($R \approx V^2/a$), 
by their {\it angular} radius, measured from a Very Long Baseline 
Interferometric (VLBI) image ($\theta_R$).  
Maser proper motions can be also be used to measure distance, but
generally yield less accurate distances than using accelerations.
Observations with the
VLBA of the \hho\ masers in \NGC\ have been carefully modeled, 
yielding the most accurate distance ($D_\theta = 7.2 \pm 0.5$ Mpc) to date 
for a galaxy \citep{Herrnstein:99}.

Unfortunately, \NGC\  is too close to determine \Ho\ directly 
(\ie by dividing its recessional velocity of 475~\kms\ by its distance),
since the galaxy's deviation from the Hubble-flow could be a significant
fraction of its recessional velocity.  Instead, the measured distance to 
\NGC\ has been used to anchor the zero point of the Cepheid 
period-luminosity relation \citep{Newman:01,Macri:06,Argon:07,Humphreys:08}.
However, galaxies with edge-on, disk-like \hho\ masers, similar to
those in \NGC, that are distant enough to be in the Hubble flow 
($D > 30$~ Mpc) could be used to measure \Ho\ directly \citep{Greenhill:04}.

Surveys of galaxies for nuclear \hho\ masers have been quite successful
and have identified more than 100 extragalactic nuclear \hho\ masers
\citep{Claussen:86,Braatz:96,Greenhill:02,Henkel:02,Greenhill:03,
Braatz:04,Kondratko:06a,Kondratko:06b,Braatz:08a}.  
In order to coordinate 
efforts to find and image new sources of nuclear \hho\ masers, we formed 
a team of scientists active in this area of research 
from the Harvard-Smithsonian Center for Astrophysics, the National Radio 
Astronomy Observatory (NRAO), and the Max-Planck-Institut f\"ur 
Radioastronomie (MPIfR).  This effort, called the Megamaser Cosmology
Project (MCP), is aimed at measuring \Ho\ directly, with $\approx 3$\%
accuracy, using a combination of VLBI imaging and single-dish monitoring
of nuclear \hho\ masers toward $\sim10$ galaxies.

Recently \citet{Braatz:08a} discovered a relatively
strong \hho\ maser ($S_\nu \approx 0.1$~Jy) toward the Seyfert 2 
nucleus of \UGC.  The \hho\ maser spectrum has the characteristics of an 
edge-on disk similar to \NGC.  The \UGC\ masers span $\approx1500$~\kms\ 
in Doppler shift and the systemic masers were observed to accelerate
by up to 8.1~\kmsperyr, suggesting an origin in a sub-pc 
disk about a $\sim10^7$~\Msun\ black hole.

In this paper, and in \citet{Braatz:08b} (hereafter Paper II), we report 
results leading to the first MCP measurement of \Ho.  
Sensitive VLBI observations and 
images of the nuclear \hho\ masers toward \UGC\ are presented in this 
paper.  In Paper II, we present monitoring observations with 
large single-dish telescopes, 
which yield accelerations of \hho\ masers.  
The combination of the VLBI imaging and the single-dish acceleration data
may yield a measurement of \Ho\ with an accuracy comparable to that of
the Hubble Key Project. 
 
\section{Observations, Calibration, and Imaging}

We observed \UGC\ on 2006 December 10 for a total of 12 hours, with
the 10 NRAO
\footnote{The National Radio Astronomy Observatory is operated by
Associated Universities, Inc., under a cooperative agreement with
the National Science Foundation.} 
VLBA antennas (under program BB227A), augmented by the Green Bank 
Telescope (GBT) and the Effelsberg 100-m telescope
\footnote{The Effelsberg 100-m telescope is a facility of the  
Max-Planck-Institut f\"ur Radioastronomie}.
The coordinates of the sources observed are listed in 
Table~\ref{table:positions}.
We alternated between two observing modes: (1) a 60-min block
of continuous tracking of \UGC\ (self-calibration mode) and (2) 
a 45-min block of rapid switching between \UGC\ and a nearby compact 
continuum source J0728+5907 (phase-referencing mode).
The phase-referencing blocks were a ``back-up'' in the event that the 
\UGC\ maser signal was not strong enough for self-calibration.
Both observing modes were successful.  However, since the 
self-calibration mode produced a much higher on-source duty cycle 
and better phase calibration than the phase-referencing mode, 
we only report results from the total of $\approx5$~hr
of self-calibration mode observations.

\begin{deluxetable}{lll}
\tablecolumns{3} \tablewidth{0pc} 
\tablecaption{Source Positions}
\tablehead {
  \colhead{Source} & \colhead{R.A. (J2000)} &  \colhead{Dec. (J2000)} 
\\
  \colhead{}       & \colhead{(h~~m~~s)} &  \colhead{(d~~'~~'')} 
            }
\startdata
 \UGC\ ...........    &07 19 30.9566  &59 21 18.330 \\
  J0728+5907 .........&07 28 47.2170  &59 07 34.128 \\
  J0753+5352 .........&07 53 01.3846  &53 52 59.637 \\
\enddata
\tablecomments {Positions used for data correlation.  
Imaging the \UGC\ maser spot at $\Vlsr=2689$~\kms, by phase-referencing
to J0728+5907, we found the maser spot
offset by ($-1,-15$)~mas (east,north) from the correlation position.}
\label{table:positions}
\end{deluxetable}

With a maximum recording rate of 512~Mbits~s$^{-1}$, we could 
cover the entire range of detectable \UGC\ \hho\ maser emission,
but not with dual-polarization for all frequency bands.
We centered 16-MHz bands at LSR velocities (optical definition) 
as follows: left circularly polarized (LCP) bands at 
3880, 3710, 3265, 2670 and 2500~\kms and
right circularly polarized (RCP) at 3880, 3265 and 2670~\kms.
The signals were sampled at the Nyquist rate (32~Mbits~s$^{-1}$) and
with 2 bits per sample.

We placed ``geodetic'' blocks at the start and end of our observations, 
in order to solve for atmospheric and clock delay residuals for each 
antenna \citep{RB04}.  In these
blocks we rapidly cycled among 14~compact radio sources that
spanned a wide range of zenith angles at all antennas.
These data were taken in left circular polarization 
with eight 16-MHz bands that spanned 492 MHz of bandwidth between 
22.00 and 22.49 GHz; the bands were spaced in a 
``minimum redundancy'' manner to sample, 
as uniformly as possible, all frequency differences.  The data were
correlated, corrected for ionospheric delays using total electron 
content measurements \citep{walker-chatterjee:00}, and residual 
multi-band delays and fringe rates were determined for all sources.  
The multi-band delays and fringe rates were modeled as owing 
to a vertical atmospheric delay and delay-rate, as well as a clock
offset and clock drift rate, at each antenna.  
Using a least-squares fitting program, we estimated zenith atmospheric 
path-delays and clock errors with accuracies typically $\approx0.5$~cm
and $\approx0.03$~nsec, respectively.

We observed the strong continuum source, J0753+5352,
hourly in order to monitor delay and electronic phase differences 
among and across the IF bands.  Generally, variations of phase across 
the VLBA bandpasses are small ($<5^\circ$) across the central 90\% of 
the band and thus we needed no bandpass corrections.  We tested the 
effect of bandpass corrections, using the J0753+5352 data,
and found position differences of $\approx0.002$~mas for maser
features mid-way between the band center and band edge. 

The raw data recorded at each antenna were 
cross-correlated with an integration time of 1.05~sec at
the VLBA correlation facility in Socorro, NM.  For this short
integration time we had to correlate the data in two passes in 
order to achieve sufficient spectral resolution (128 spectral 
channels for each IF band) without exceeding the maximum correlator 
output rate.  Before calibration, the two correlation data sets 
were ``glued'' together.

We calibrated the data using the NRAO Astronomical Image 
Processing System (AIPS).  
First, we corrected interferometer delays and phases for
the effects of diurnal feed rotation (parallactic angle) and
for small errors in the values of the Earth's orientation parameters
used at the time of correlation.  By analyzing the data taken in
phase-referencing mode, we determined that the strong maser
feature at $\Vlsr=2689$~\kms, which we later used as the phase
reference for the self-calibration mode data, was offset 
from the position of \UGC\ used in the VLBA correlator
by ($-1,-15$)~mas toward (east,north), respectively, relative 
to J0728+5907.  

Since the VLBA correlator model includes no ionospheric delays, we used 
global total electron content models to remove ionospheric effects.
We then corrected the data for residual zenith atmospheric 
delays and clock drifts, as determined from the geodetic block data.  
While we obtained good atmospheric/clock corrections for most
antennas, insufficient data were obtained for the Effelsberg (EB) and
Mauna Kea (MK) antennas for this task.  Thus, we later used
the data from the hourly observations of J0753+5352 to determine
final delay corrections for the \UGC\ data.

We corrected the interferometer visibility amplitudes for 
the few percent effects of biases in the threshold levels of the data
samplers at each antenna.  We also entered system temperature and
antenna gain curve information into calibration tables. 
These tables were used later to convert correlation coefficients 
to flux densities.
Next, we performed a ``manual phase-calibration'' to remove
delay and phase differences among all bands.  This was
accomplished with data from one scan on a strong calibrator, 4C~39.25.
We did not shift the frequency axes of the maser interferometer spectra
to compensate for the Doppler shift changes during the $\pm5$~hr
\UGC\ observing track, as these effects were less than our velocity 
resolution of 1.7~\kms.

The final calibration involved selecting a maser feature
as the interferometer phase-reference.  The strongest maser
feature in the spectrum peaked at $\approx0.07$~Jy and was fairly
broad.  We found that using 5 channels spanning an LSR velocity 
range of 2685 to 2692~\kms\ (\ie\ channels 52 to 56 from the blue-shifted
high velocity band centered at $\Vlsr=2670$~\kms), adding
together both polarizations, and fitting fringes over a 1~min period 
gave optimum results.  The St. Croix (SC) antenna failed to produce
phase-reference solutions and data from that antenna were discarded.
For most antennas at most times the phases could be easily interpolated 
between solutions.  However, when the differences
between adjacent reference phases exceeded $60^\circ$,  
the data between those times were discarded.  This editing
was done on baseline (not antenna) data, since correlated phases 
between antennas do not affect interferometer coherence.

After calibration, we Fourier transformed the gridded \uvdata\ 
to make images of the maser emission in all spectral 
channels for each of the five IF bands.  The point-source response
function had FWHM of $0.35\times0.22$~mas elongated along a position 
angle of $-17^\circ$ east of north.  The images were 
deconvolved with the point-source response using the CLEAN algorithm
and restored with a circular Gaussian beam with a 0.30~mas FWHM.
All images appeared to contain single, point-like maser spots. 
We then fitted each spectral channel image with an elliptical Gaussian 
brightness distribution in order to obtain positions and flux densities.

\section        {Results \& Discussion}

\begin{figure}
\epsscale{0.9} 
\plotone{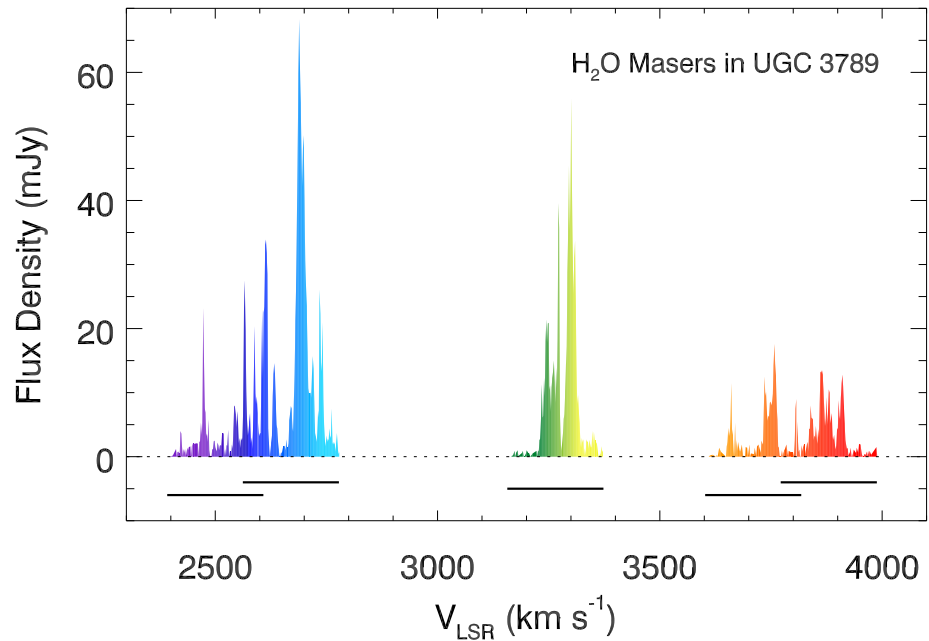} 
\caption{Interferometer spectrum of the 22~GHz \hho\ masers toward
\UGC\ constructed from VLBI data using the VLBA, the GBT and the
Effelsberg antennas.   The systemic velocity of the galaxy of
$3325\pm24$~\kms, as determined from HI observations,
is within the systemic velocity components shown in {\it green} colors.
High velocity components, shifted by up to $800$~\kms\ from the
systemic velocity, are shown in {\it blue} and {\it red} colors.  
Broad-band spectra taken with the GBT before the VLBI observations 
showed almost no detectable maser
features outside our observing bands, indicated by horizontal 
lines below the spectrum.
  \label{fig:spectrum}
        }
\end{figure}

Channel maps typically had rms noise levels of $\approx0.9$~mJy for the
dual-polarized IF bands and $\approx1.2$~mJy for the single-polarization
IF bands.  The flux densities from the Gaussian fits for all spectral 
channels in all IF bands were used to generate the interferometer spectrum 
shown in Fig.~\ref{fig:spectrum}.  When little signal was detected in a 
spectral channel, as evidenced by a failed fit or a spot size greater 
than 1~mas, we assigned that channel zero flux density.

\begin{figure}
\epsscale{0.9} 
\plotone{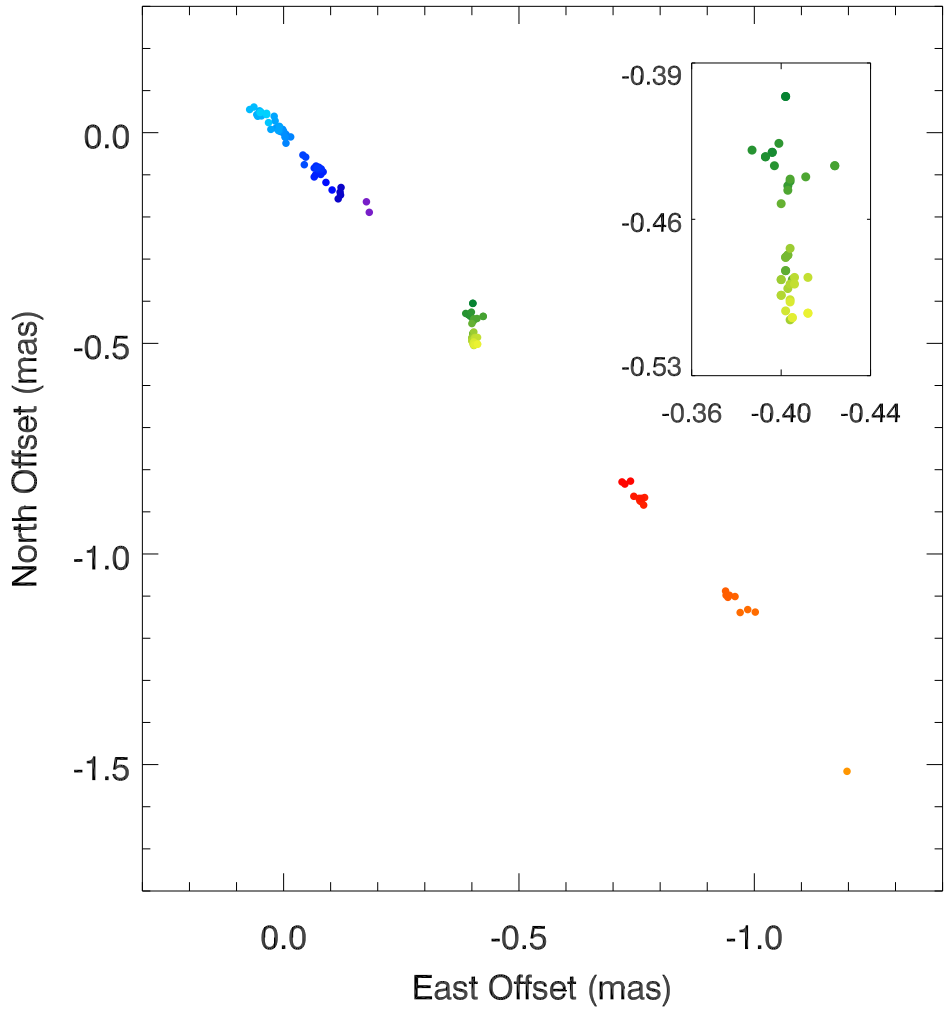}
\caption{Map of the relative positions of individual maser spots toward \UGC.
High velocity blue-shifted ({\it blue}) and red-shifted ({\it red}) masers
straddle the systemic masers ({\it green} and expanded view {\it inset}) 
and the linear
arrangement of spots suggests that we are viewing a nearly edge-on
rotating disk, similar to that seen in \NGC.  The $\approx2$~mas extent
of the maser spots in \UGC\ is approximately seven times smaller than for
\NGC, which is consistent with \UGC\ being at approximately seven times 
greater distance.  Formal fitting 
uncertainties are given in Tables~\ref{table:reds}, \ref{table:systemics}
and \ref{table:blues} and are typically $<0.010$~mas.  
  \label{fig:spots}
        }
\end{figure}

The flux densities and positions determined by Gaussian fitting each 
spectral channel image are reported in Tables~\ref{table:reds},
\ref{table:systemics} and \ref{table:blues} for maser spots stronger 
than 10~mJy.  The positions of these spots are plotted in 
Fig.~\ref{fig:spots}.   
The nearly linear arrangement of the maser spots on the sky is striking.  
The red- and blue-shifted high-velocity spots straddle the
systemic emission complex.   This spatial-velocity arrangement 
is characteristic of a nearly edge-on disk, as is well documented
for \NGC\ \citep{Herrnstein:05}.

\begin{figure}
\epsscale{0.9} 
\plotone{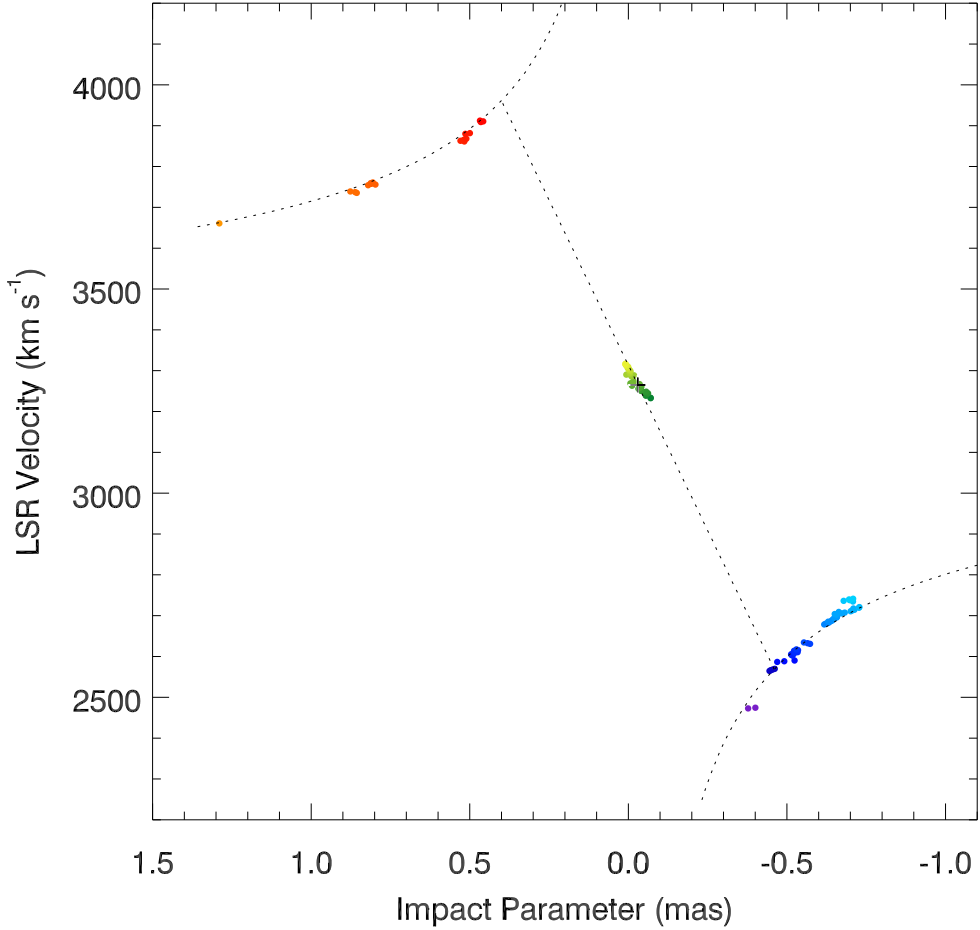}
\caption{Position-Velocity plot of the maser spots toward \UGC.
The high velocity blue-shifted ({\it blue}) and red-shifted
({\it red}) spots display a Keplerian $1/\sqrt{R}$ rotation
curve, indicated by the {\it curved dotted lines}.
The systemic masers ({\it green}) are consistent with projected positions
and velocities for gas in Keplerian orbit at $R\approx0.43$~mas,
indicated by the {\it straight dotted line},
but small deviations from a linear distribution are apparent.
Impact parameter is defined as distance along a position angle of
$41^\circ$ east of north from an (east,north) offset of ($-0.4,-0.5$)~mas; 
the plus sign ($+$) at ($-0.03$ mas, $3265$ \kms) indicates the assumed 
center of the SMBH.
  \label{fig:pv}
        }
\end{figure}

We calculated the position along the spot distribution (\ie an impact 
parameter along position angle of $41^\circ$ east of north)
and show a position-velocity plot in Fig.~\ref{fig:pv}.
The high-velocity masers display a Keplerian velocity ($V\propto1/\sqrt{R}$)
versus impact parameter (or radius), suggesting that the gravitational
potential is dominated by a SMBH.  The Keplerian velocity pattern is 
centered at $\Vlsr \approx 3265$~\kms.   This is slightly offset 
from the central velocity of HI emission from the galaxy at 
$\Vhelio \approx 3325\pm24$~\kms\ \citep{Theureau:98}.
(Note: $\Vlsr - V_{\rm Helio} = 0.3$~\kms\ for \UGC.) 
Correcting the maser velocity to the CMB reference frame (\ie
$V_{\rm CMB} \approx \Vlsr + 60~\kms$), yields a recessional velocity 
of 3325~\kms.   Thus, for $\Ho=72$~\kms~Mpc$^{-1}$, 
\UGC's distance would be expected to be $\approx46$ Mpc.

The detected blue-shifted high velocity masers sample disk radii
between 0.35 and 0.70~mas (0.08 to 0.16~pc) and achieve 
rotation speeds as high as 792~\kms, with respect to a central systemic 
velocity of 3265~\kms.  The detected red-shifted masers sample radii of 
0.50 to 1.33~mas (0.11 to 0.30~pc) and achieve rotation speeds up to 
647~\kms.  Also shown by the straight dotted lines in Fig.~\ref{fig:pv} 
is the position-velocity distribution expected for systemic maser spots 
that lie at a radius of 0.43~mas from the central mass, whose assumed
location is indicated by the plus-sign ($+$) in the figure.  
These spatial-kinematic parameters are comparable to those of the
\hho\ masers in \NGC, which sample radii of $\approx0.14$ to $0.28$~pc 
and achieve rotation speeds of $\approx1100$~\kms.  The moderately lower 
rotation speed at a slightly smaller radius suggests that the SMBH
at the center of \UGC\ is less massive than the $3.9\times10^7$~\Msun\ 
SMBH in \NGC.  At a distance of 46~Mpc, the high-velocity
data for \UGC\ can be well fit by gas in circular orbit about a central 
mass of $1.1\times10^7$~\Msun, as shown by the blue and red dotted lines in 
Fig.~\ref{fig:pv}.   

As can be seen in Fig.~\ref{fig:spots}, the systemic features 
lie between the high velocity features but are distributed along
a position angle of roughly $10^\circ$ (east of north).  Thus, they are
misaligned by approximately $30^\circ$ with respect to the $41^\circ$
position angle of the disk, obtained by drawing a line through
the high velocity masers.  This suggests that the \UGC\ disk may be
slightly inclined and warped and/or that the systemic masers
are not all at the same radius.  (Note that the \NGC\ disk is
both inclined by $8^\circ$ with respect to our line of sight and
warped \citep{Herrnstein:05}.)  Deviation from a perfectly flat,
edge-on disk for \UGC\ can also be seen in the position-velocity
plot (Fig.~\ref{fig:pv}) as a slight bending of the systemic maser 
spots and in the variation in accelerations seen by \citet{Braatz:08a}.
Modeling of the disk will need to accommodate these complications.

We searched for continuum emission from the vicinity of the SMBH
(\ie\ near the position of the systemic velocity masers) by summing
channels 5 through 120 of the (red shifted) dual-circularly polarized
band centered at $\Vlsr=3380$~\kms.   We maximized the detection 
sensitivity by natural weighting of the data when imaging.
The masers were detected at the expected position, offset by 
$\approx1$~mas from the position of the SMBH, but no continuum 
emission was detected at a $2\sigma$ limit of $<0.14$~mJy.

\section        {Conclusions}

The discovery of \hho\ masers emanating from a sub-pc disk 
in the Seyfert 2 galaxy \NGC\ more than two decades 
ago has led to detailed imaging of an AGN accretion disk.
Geometric modeling of the Keplerian orbits of the masers 
yielded the most accurate distance to any galaxy, allowing
recalibration of the extragalactic distance scale.  Now,
the recent discovery by \citet{Braatz:08a} of \hho\ masers in 
\UGC\ offers the opportunity to extend this technique to a
galaxy seven times more distant.   

In this paper we presented VLBI images of the \UGC\ \hho\ masers, 
which showed that these masers are remarkably similar to those in \NGC.
In both sources, the spatial distribution is nearly linear, with high 
velocity masers on both sides (both spatially and spectrally) of 
systemic velocity masers.  The masers trace gas in Keplerian orbits 
with rotation speeds of $\sim1000$~\kms\ at radii of $\sim0.1$~pc, 
presumably moving under the influence of a $\sim10^7$~\Msun\ SMBH.

\UGC\ has a recessional velocity of $\approx 3325$~\kms\ and is well 
into the Hubble flow.  The VLBI results presented in this paper will 
be followed by detailed spectral monitoring data and disk modeling 
in Paper II to determine the distance to \UGC.  This angular-diameter 
distance, when combined with its recessional velocity, 
should yield a direct and accurate estimate of \Ho.


\vskip 0.5truecm
{\it Facilities:} \facility{VLBA, GBT, Effelsberg}

\begin{deluxetable}{ccccrrrr}
\tablecolumns{8} 
\tablewidth{0pc} 
\tablecaption{\UGC\ Red-shifted High-velocity Spots}
\tablehead {
  \colhead{\Vlsr} & \colhead{Flux Density} &&  
  \colhead{$\Theta_x$} & \colhead{$\sigma_{\Theta_x}$} &&
  \colhead{$\Theta_y$} & \colhead{$\sigma_{\Theta_y}$} 
\\
  \colhead{(\kms)}  & \colhead{(mJy)} &&  
  \colhead{(mas)} & \colhead{(mas)} &&
  \colhead{(mas)} & \colhead{(mas)} 
            }
\startdata
    3912.3.....&  11.5 &&$-$0.737 &0.009  &&$-$0.827 &0.013 \cr
    3910.6.....&  13.3 &&$-$0.719 &0.008  &&$-$0.829 &0.012 \cr
    3908.9.....&  12.0 &&$-$0.725 &0.009  &&$-$0.834 &0.013 \cr
    3881.7.....&  10.5 &&$-$0.744 &0.010  &&$-$0.863 &0.015 \cr
    3880.0.....&  11.0 &&$-$0.760 &0.010  &&$-$0.868 &0.014 \cr
    3868.1.....&  10.8 &&$-$0.755 &0.010  &&$-$0.868 &0.014 \cr
    3866.4.....&  13.3 &&$-$0.757 &0.008  &&$-$0.874 &0.012 \cr
    3864.7.....&  14.0 &&$-$0.760 &0.008  &&$-$0.876 &0.011 \cr
    3863.0.....&  13.6 &&$-$0.765 &0.008  &&$-$0.884 &0.011 \cr
    3861.3.....&  13.8 &&$-$0.767 &0.008  &&$-$0.866 &0.011 \cr
\cr
    3761.0.....&  13.2 &&$-$0.940 &0.011  &&$-$1.098 &0.015 \cr
    3759.3.....&  15.0 &&$-$0.944 &0.009  &&$-$1.103 &0.014 \cr
    3757.6.....&  18.1 &&$-$0.947 &0.008  &&$-$1.098 &0.011 \cr
    3755.9.....&  14.8 &&$-$0.939 &0.009  &&$-$1.088 &0.014 \cr
    3754.2.....&  10.3 &&$-$0.959 &0.014  &&$-$1.101 &0.020 \cr
    3738.9.....&  10.6 &&$-$1.002 &0.013  &&$-$1.138 &0.019 \cr
    3737.2.....&  10.0 &&$-$0.986 &0.014  &&$-$1.132 &0.020 \cr
    3735.5.....&  13.0 &&$-$0.970 &0.011  &&$-$1.139 &0.016 \cr
    3660.7.....&  12.0 &&$-$1.197 &0.012  &&$-$1.516 &0.017 \cr
\enddata
\tablecomments {Columns 1 and 2 give the LSR velocity and flux
density of maser spots in individual spectral channels.  
Columns 3 (5) and 4 (6) give the east (north) offsets and their 
uncertainties.
Offsets are with respect to the phase reference obtained by
summing the emission between velocities 2685 and 2692~\kms.}
\label{table:reds}
\end{deluxetable}

\begin{deluxetable}{ccccrrrr}
\tablecolumns{8} 
\tablewidth{0pc} 
\tablecaption{\UGC\ Systemic Velocity Spots}
\tablehead {
  \colhead{\Vlsr} & \colhead{Flux Density} &&  
  \colhead{$\Theta_x$} & \colhead{$\sigma_{\Theta_x}$} &&
  \colhead{$\Theta_y$} & \colhead{$\sigma_{\Theta_y}$} 
\\
  \colhead{(\kms)}  & \colhead{(mJy)} &&  
  \colhead{(mas)} & \colhead{(mas)} &&
  \colhead{(mas)} & \colhead{(mas)} 
            }
\startdata
    3316.0.....&  10.1 &&$-$0.412 &0.010  &&$-$0.502 &0.015 \cr
    3312.6.....&  13.7 &&$-$0.405 &0.008  &&$-$0.504 &0.011 \cr
    3310.9.....&  24.1 &&$-$0.413 &0.004  &&$-$0.488 &0.006 \cr
    3309.2.....&  34.2 &&$-$0.404 &0.003  &&$-$0.497 &0.004 \cr
    3307.5.....&  31.6 &&$-$0.409 &0.003  &&$-$0.498 &0.005 \cr
    3305.8.....&  25.9 &&$-$0.402 &0.004  &&$-$0.501 &0.006 \cr
    3304.1.....&  31.1 &&$-$0.404 &0.003  &&$-$0.496 &0.005 \cr
    3302.4.....&  42.3 &&$-$0.412 &0.002  &&$-$0.486 &0.004 \cr
    3300.7.....&  56.4 &&$-$0.406 &0.002  &&$-$0.489 &0.003 \cr
    3299.0.....&  41.8 &&$-$0.406 &0.003  &&$-$0.486 &0.004 \cr
    3297.3.....&  35.7 &&$-$0.404 &0.003  &&$-$0.489 &0.004 \cr
    3295.6.....&  45.9 &&$-$0.400 &0.002  &&$-$0.494 &0.003 \cr
    3293.9.....&  36.4 &&$-$0.402 &0.003  &&$-$0.483 &0.004 \cr
    3292.2.....&  29.4 &&$-$0.403 &0.004  &&$-$0.491 &0.005 \cr
    3290.5.....&  17.7 &&$-$0.404 &0.006  &&$-$0.505 &0.009 \cr
    3288.8.....&  14.9 &&$-$0.404 &0.007  &&$-$0.473 &0.010 \cr
    3287.1.....&  10.8 &&$-$0.400 &0.010  &&$-$0.487 &0.014 \cr
    3273.5.....&  36.2 &&$-$0.403 &0.003  &&$-$0.476 &0.004 \cr
    3271.8.....&  40.1 &&$-$0.402 &0.003  &&$-$0.477 &0.004 \cr
    3270.1.....&  18.5 &&$-$0.398 &0.006  &&$-$0.477 &0.008 \cr
    3268.4.....&  11.2 &&$-$0.405 &0.009  &&$-$0.487 &0.014 \cr
    3266.7.....&  14.4 &&$-$0.400 &0.007  &&$-$0.453 &0.011 \cr
    3263.3.....&  12.2 &&$-$0.402 &0.009  &&$-$0.483 &0.013 \cr
    3261.6.....&  15.5 &&$-$0.403 &0.007  &&$-$0.447 &0.010 \cr
    3259.9.....&  13.9 &&$-$0.404 &0.008  &&$-$0.442 &0.011 \cr
    3258.2.....&  12.8 &&$-$0.411 &0.008  &&$-$0.441 &0.012 \cr
    3256.5.....&  11.3 &&$-$0.424 &0.009  &&$-$0.436 &0.014 \cr
    3253.1.....&  11.6 &&$-$0.404 &0.009  &&$-$0.443 &0.013 \cr
    3251.4.....&  10.9 &&$-$0.403 &0.010  &&$-$0.447 &0.014 \cr
    3249.7.....&  21.4 &&$-$0.403 &0.005  &&$-$0.445 &0.007 \cr
    3248.0.....&  21.3 &&$-$0.399 &0.005  &&$-$0.426 &0.007 \cr
    3246.3.....&  18.3 &&$-$0.397 &0.006  &&$-$0.436 &0.008 \cr
    3244.6.....&  21.9 &&$-$0.387 &0.005  &&$-$0.429 &0.007 \cr
    3242.9.....&  18.6 &&$-$0.393 &0.006  &&$-$0.432 &0.008 \cr
    3239.5.....&  12.8 &&$-$0.396 &0.008  &&$-$0.430 &0.012 \cr
    3234.4.....&  11.8 &&$-$0.402 &0.009  &&$-$0.405 &0.013 \cr
\enddata
\tablecomments {See Table~\ref{table:reds}} 
\label{table:systemics}
\end{deluxetable}

\begin{deluxetable}{cccrrrrr}
\tablecolumns{8} 
\tablewidth{0pc} 
\tablecaption{\UGC\ Blue-shifted High-velocity Spots}
\tablehead {
  \colhead{\Vlsr} & \colhead{Flux Density} &&  
  \colhead{$\Theta_x$} & \colhead{$\sigma_{\Theta_x}$} &&
  \colhead{$\Theta_y$} & \colhead{$\sigma_{\Theta_y}$} 
\\
  \colhead{(\kms)}  & \colhead{(mJy)} &&  
  \colhead{(mas)} & \colhead{(mas)} &&
  \colhead{(mas)} & \colhead{(mas)} 
            }
\startdata
    2741.4.....&  21.5 && 0.050 &0.005  && 0.048 &0.007 \cr
    2739.7.....&  15.2 && 0.036 &0.007  && 0.043 &0.010 \cr
    2738.0.....&  13.8 && 0.036 &0.008  && 0.046 &0.011 \cr
    2736.3.....&  22.8 && 0.032 &0.005  && 0.024 &0.007 \cr
    2734.6.....&  26.6 && 0.048 &0.004  && 0.049 &0.006 \cr
    2721.0.....&  14.2 && 0.072 &0.007  && 0.055 &0.011 \cr
    2719.3.....&  16.2 && 0.063 &0.006  && 0.061 &0.009 \cr
    2717.6.....&  13.4 && 0.057 &0.008  && 0.043 &0.011 \cr
    2714.2.....&  10.5 && 0.051 &0.010  && 0.052 &0.015 \cr
    2712.5.....&  10.4 && 0.055 &0.010  && 0.039 &0.015 \cr
    2710.8.....&  10.4 && 0.047 &0.010  && 0.040 &0.015 \cr
    2709.1.....&  11.2 && 0.027 &0.009  && 0.008 &0.014 \cr
    2707.4.....&  16.9 && 0.020 &0.006  && 0.039 &0.009 \cr
    2705.7.....&  24.1 && 0.018 &0.004  && 0.028 &0.006 \cr
    2704.0.....&  27.3 && 0.011 &0.004  && 0.005 &0.006 \cr
    2702.3.....&  31.1 && 0.009 &0.003  && 0.014 &0.005 \cr
    2700.6.....&  40.7 && 0.014 &0.003  && 0.012 &0.004 \cr
    2698.9.....&  50.7 && 0.016 &0.002  && 0.011 &0.003 \cr
    2697.2.....&  48.7 && 0.014 &0.002  && 0.011 &0.003 \cr
    2695.5.....&  42.7 && 0.009 &0.002  && 0.015 &0.004 \cr
    2693.8.....&  48.5 && 0.008 &0.002  && 0.006 &0.003 \cr
    2692.1.....&  55.8 && 0.002 &0.002  && 0.008 &0.003 \cr
    2690.4.....&  61.3 && 0.007 &0.002  && 0.003 &0.003 \cr
    2688.7.....&  68.8 && 0.000 &0.002  && 0.003 &0.002 \cr
    2687.0.....&  60.4 && 0.001 &0.002  && 0.000 &0.003 \cr
    2685.3.....&  48.4 &&$-$0.003 &0.002  &&$-$0.010 &0.003 \cr
    2683.6.....&  37.5 &&$-$0.005 &0.003  &&$-$0.004 &0.004 \cr
    2681.9.....&  27.5 &&$-$0.004 &0.004  &&$-$0.011 &0.006 \cr
    2680.2.....&  21.8 &&$-$0.015 &0.005  &&$-$0.010 &0.007 \cr
    2678.5.....&  13.3 &&$-$0.005 &0.008  &&$-$0.025 &0.012 \cr
    2634.3.....&  13.5 &&$-$0.044 &0.008  &&$-$0.076 &0.011 \cr
    2632.6.....&  15.1 &&$-$0.047 &0.007  &&$-$0.058 &0.010 \cr
    2630.9.....&  13.4 &&$-$0.041 &0.008  &&$-$0.053 &0.011 \cr
\tablebreak
    2617.3.....&  29.0 &&$-$0.071 &0.004  &&$-$0.084 &0.005 \cr
    2615.6.....&  32.3 &&$-$0.069 &0.003  &&$-$0.079 &0.005 \cr
    2613.9.....&  34.4 &&$-$0.080 &0.003  &&$-$0.086 &0.004 \cr
    2612.2.....&  33.6 &&$-$0.075 &0.003  &&$-$0.082 &0.005 \cr
    2610.5.....&  23.9 &&$-$0.065 &0.004  &&$-$0.084 &0.006 \cr
    2608.8.....&  23.0 &&$-$0.074 &0.005  &&$-$0.088 &0.007 \cr
    2607.1.....&  12.6 &&$-$0.068 &0.008  &&$-$0.100 &0.012 \cr
    2605.4.....&  23.5 &&$-$0.079 &0.004  &&$-$0.099 &0.007 \cr
    2603.7.....&  13.0 &&$-$0.080 &0.008  &&$-$0.112 &0.012 \cr
\cr
    2602.0.....&  13.6 &&$-$0.065 &0.010  &&$-$0.105 &0.015 \cr
    2590.1.....&  10.4 &&$-$0.074 &0.013  &&$-$0.089 &0.020 \cr
    2588.4.....&  20.9 &&$-$0.090 &0.007  &&$-$0.118 &0.010 \cr
    2586.7.....&  12.8 &&$-$0.103 &0.011  &&$-$0.136 &0.016 \cr
    2569.7.....&  10.7 &&$-$0.122 &0.013  &&$-$0.130 &0.019 \cr
    2568.0.....&  20.2 &&$-$0.120 &0.007  &&$-$0.141 &0.010 \cr
    2566.3.....&  28.0 &&$-$0.121 &0.005  &&$-$0.148 &0.007 \cr
    2564.6.....&  23.7 &&$-$0.116 &0.006  &&$-$0.157 &0.009 \cr
    2474.5.....&  15.7 &&$-$0.176 &0.009  &&$-$0.164 &0.013 \cr
    2472.8.....&  23.7 &&$-$0.182 &0.006  &&$-$0.189 &0.009 \cr
\enddata
\tablecomments {See Table~\ref{table:reds}} 
\label{table:blues}
\end{deluxetable}


\begin{thebibliography}{}
\bibitem[Argon \etal (2007)]{Argon:07}
   Argon, A. L., Greenhill, L. J., Reid, M. J., Moran, J. M.
   \& Humphreys, E. M. L. 2007, \apj, 659, 1040
\bibitem[Braatz, Wilson \& Henkel (1996)]{Braatz:96}
   Braatz, J.~A., Wilson, A. S. \& Henkel, C. 1996, \apjs, 106, 51
\bibitem[Braatz \etal (2004)]{Braatz:04}
   Braatz, J.~A., Henkel, C., Greenhill, L., Moran, J. \& Wilson, A. 
   2004, \apj, 617, L29
\bibitem[Braatz \& Gugliucci (2008)]{Braatz:08a}
   Braatz, J. A. \& Gugliucci, N. 2008, \apj, 678, 96
\bibitem[Braatz \etal (2008)]{Braatz:08b}
   Braatz \etal 2008 (Paper II), in preparation
\bibitem[Claussen \& Lo (1986)]{Claussen:86}
   Claussen, M. J. \& Lo, K.-Y. 1986, \apj, 308, 592
\bibitem[Freedman \etal (2001)]{Freedman:01}
   Freedman, W. \etal 2001, \apj, 553, 47
\bibitem[Greenhill \etal~(1995a)]{Greenhill:95a}
   Greenhill, L. J., Jiang, D. R., Moran, J. M., Reid, M. J., 
   Lo, K. Y. \& Claussen, M. J. 1995a, \apj, 440, 619
\bibitem[Greenhill \etal~(1995b)]{Greenhill:95b}
   Greenhill, L.~J., Henkel, C., Becker, R., Wilson, T.~L. \& 
   Wouterloot, J.~G.~A. 1995b, \aap, 304, 21
\bibitem[Greenhill \etal (2002)]{Greenhill:02}
   Greenhill \etal 2002, \apj, 565, 836
\bibitem[Greenhill \etal (2003)]{Greenhill:03}
   Greenhill, L. J., Kondratko, P. T., Lovell, J. E. J., Kuiper, T. B. H., 
   Moran, J. M., Jauncey, D. L. \& Baines, G. P. 2003, \apj, 582, L11
\bibitem[Greenhill (2004)]{Greenhill:04}
   Greenhill, L. J. 2004, NewAR, 48, 1079
\bibitem[Haschick, Baan \& Peng~(1994)]{Haschick:94}
   Haschick, A. D., Baan, W. A. \& Peng, E. W. 1994, \apj, 437, L35
\bibitem[Henkel \etal (2002)]{Henkel:02}
   Henkel, C., Braatz, J. A., Greenhill, L. J. \& Wilson, A. S.
   2002, \aap, 394, L23
\bibitem[Herrnstein \etal~(1999)]{Herrnstein:99}
   Herrnstein \etal~1999, \nat, 400, 539
\bibitem[Herrnstein \etal (2005)]{Herrnstein:05}
   Herrnstein, J. R., Moran, J. M., Greenhill, L. J. and Trotter, A. S.
   2005, \apj, 629, 719
\bibitem[Hu (2005)]{Hu:05}
   Hu, W. 2005, ASP Conf Series Vol 339, Observing Dark Energy, 
   eds. S.C. Wolff \& T.R. Lauer (San Francisco: ASP), 215
\bibitem[Humphreys \etal (2008)]{Humphreys:08}
   Humphreys, E. M. L., Reid, M. J., Greenhill, L. J., Moran, J. M. 
   \& Argon, A. L. 2008, \apj, 672, 800
\bibitem[Kondratko \etal (2006)]{Kondratko:06a}
   Kondratko, P. T. \etal 2006, \apj, 638, 100
\bibitem[Kondratko, Greenhill \& Moran (2006)]{Kondratko:06b}
   Kondratko, P. T., Greenhill, L. J. \& Moran, J. M. 2006,
   \apj, 652, 136
\bibitem[Macri \etal (2006)]{Macri:06}
   Macri, L. M., Stanek, K. Z., Bersier, D., Greenhill, L. J.
   \& Reid, M. J. 2006, \apj, 652, 1133
\bibitem[Miyoshi \etal (1995)]{Miyoshi:95}
   Miyoshi, M., Moran, J., Herrnstein, J.,  Greenhill, L., Nakai, N., 
   Diamond, P. \& Inoue, M. 1995, \nat, 373, 127
\bibitem[Newman \etal (2001)]{Newman:01}
   Newman, J. A. \etal 2001, \apj, 553, 562
\bibitem[Ratra \& Peebles (1988)]{Ratra:88}
   Ratra, B. \& Peebles, P.~J.~E. 1988, Phys. Rev. D, 37, 3406
\bibitem[Reid \& Brunthaler (2004)]{RB04}
   Reid, M. J. \& Brunthaler, A. 2004, \apj, 616, 872
\bibitem[Spergel \etal (2003)]{Spergel:03} 
   Spergel, D. N. \etal 2003, \apjs, 148, 175
\bibitem[Theureau \etal (1998)]{Theureau:98}
   Theureau \etal\ 1998, \aaps, 130, 333
\bibitem[Walker \& Chatterjee (2000)]{walker-chatterjee:00}
    Walker, C. \& Chatterjee, S., 2000, VLBA Scientific Memo 23,
    http://www.nrao.edu/memos/sci/gps\_ion.html
\bibitem[Watson \& Wallin~(1994)]{Watson:94}
   Watson, W. \& Wallin, B. 1994, \apj, 432, L35
\bibitem[Wetterich (1988)]{Wetterich:88}
   Wetterich, C. 1988, Nucl. Phys. B, 302, 668
\end{thebibliography}
\end{document}